\documentclass[%
reprint,
showpacs,preprintnumbers,
 amsmath,amssymb,
 aps,
]{revtex4-1}

\usepackage{graphicx}% Include figure files
\usepackage{dcolumn}% Align table columns on decimal point
\usepackage{bm}% bold math
\usepackage{subfigure} % Written by Steven Douglas Cochran
\usepackage{color,soul}
\usepackage{ulem}
\usepackage{booktabs}
\usepackage{amsmath}
\usepackage{ulem}
\begin{document}

%\preprint{APS/2020}

\title{Free-energy landscapes of intrusion and extrusion of liquid in truncated and inverted truncated conical pores: implications to the Cassie-Baxter to Wenzel transition% Force line breaks with \\
}

\author{Masao Iwamatsu}
\affiliation{
Tokyo City University, Setagaya-ku, Tokyo 158-8557, Japan
}
%\altaffiliation[Also at ]{Department of Physics, Tokyo City University, Setagaya-ku, Tokyo 158-8557, Japan}%Lines %break automatically or can be forced with \\
\email{iwamatm@tcu.ac.jp}
%\author{Hiroyuki Mori}%
%\affiliation{
%Department of Physics, Tokyo Metropolitan University, Hachioji, Tokyo 192-0397, Japan

%\affiliation{
%\affiliation{
%Third institution, the second for Charlie Author
%\affiliation{%


\date{\today}% It is always \today, today,

\begin{abstract}
As the simplest model of transition between the superhydrophobic Cassie-Baxter (CB) and Wenzel (W) states of a macroscopic droplet sitting on a microscopically rough or corrugated substrate, a substrate whose surface is covered by identical truncated or inverted truncated conical pores is considered. The free energy landscapes of the intrusion and extrusion processes of a liquid into single pore are analyzed when the liquid is compressed or stretched so that the liquid phase is either stable or metastable relative to the vapor phase. Therefore, this model is also relevant to the stability of the superhydrophobic submerged substrates. In this study, the macroscopic classical capillary theory is adopted. Even within this simplified model, two simple geometries of truncated and inverted truncated cones lead to completely different free-energy landscapes. A simple criterion for the stability of the CB state based on Laplace pressure is shown not to be sufficient to understand the destruction and recovery of the CB state. The free-energy landscapes indicate that a gradual and an abrupt destruction of CB state is possible, which depends on the orientation of the conical pore and whether the liquid is compressed or stretched.  The extensions of these theoretical results to more complex geometries are briefly discussed.
\end{abstract}

%\pacs{64.60.Q-}% PACS, the Physics and Astronomy
%Use showkeys class option if keyword
\maketitle

\section{Introduction}

Intrusion and extrusion of a liquid in microscale and nanoscale pores is a fundamental problem of thermodynamics of liquid in confined space~\cite{Debenedetti1996,Kelton2010,Lefevre2004,Guillemont2012,Tinti2017}.  This problem is related to the wetting and drying of a small pore, and heterogeneous nucleation from a metastable liquid or vapor phase.  Recently, there has been growing interests in addressing this problem because it is relevant to  the engineering~\cite{Verho2012,Bormashenko2015,Xue2016,Jiang2020} of the so called superhydrophobic (SHP) substrates, which are realized from the Cassie-Baxter (CB) state~\cite{Cassie1944} of a surface.  To fabricate an SHP Cassie-Baxter (SHP-CB) state, pores of various shapes randomly or regularly distributed are engraved artificially on the surface of the substrate to make the substrate rough.  If these pores are completely dry and filled with vapor, the liquid on the surface is supported by the cushion formed by the vapor insides the pores.  Furthermore, the substrate shows superhydrophobicity which is characterized by a contact angle larger than $150^{\circ}$ and a small contact angle hysteresis~\cite{Verho2012,Bormashenko2015,Xue2016,Jiang2020}

The superhydrophobicity exhibited by the substrate can deteriorate due to the wetting transition in a pore from a completely dry CB state to a completely wet Wenzel (W)~\cite{Wenzel1936} state~\cite{Verho2012,Bormashenko2015,Xue2016,Jiang2020}.   In the W state, all the pores are completely wet and filled with liquid.  Therefore, preventing the wetting transition of the pore is crucial to achieve a sustainable superhydrophbicity.  For this reason, the CB  to W wetting transition has attracted considerable attention for the last two decades.   Moreover, the W to CB drying transition is important to understand the restoration of the SHP-CB state~\cite{Verho2012,Checco2014,Singh2015,Amabili2016,Prakash2016,Jones2017,Giacomello2019}.

Early theoretical studies~\cite{Marmur2003,Marmur2006} were based on the classical Cassie-Baxter model~\cite{Cassie1944} and the Wenzel model~\cite{Wenzel1936} of the apparent contact angle.  In these models, the wettability represented by the intrinsic Young's contact angle and the solid-liquid and solid-vapor surface areas, determines the apparent contact angle, which can be more hydrophobic or hydrophilic compared to the original Young's contact angle.   Later, not only the surface area but also the geometrical shape of the pore is found to be important~\cite{Tuteja2007,Nosonovsky2007,Verho2012,Bormashenko2015,Amabili2015,Xue2016,Jiang2020}.  The SHP-CB state can be stabilized by pinning a liquid-vapor interface (edge effect) at the inlet of a pore and it can be destroyed by the sagging mechanism~\cite{Patanker2010,Papadopoulos2013}. However, the simple intrusion and extrusion of a liquid into a pore is found to be the most basic process of the transition between the CB and W states
~\cite{Verho2012,Checco2014,Singh2015,Amabili2016,Prakash2016,Jones2017,Giacomello2019}.  Therefore, the knowledge of the free-energy landscape and the energy barrier~\cite{Verho2012,Checco2014,Singh2015,Amabili2015,Amabili2016,Prakash2016,Jones2017,Giacomello2019,Whyman2011,Savoy2012,
Giacomello2012,Bormashenko2013,Iwamatsu2016}, which separates both the CB and W states, would be crucial to understand the stability and recovery of the SHP-CB state.

So far, we have used the Cassie-Baxter (CB) state and the superhydrophobic Cassie-Baxter (SHP-CB) state interchangeably.  From now on, we will use the CB state to represent the completely dry single pore and the SHP-CB state to represent the superhydrophobic state of substrate to distinguish between the wetting transition of the individual pore and that of the substrate.

In this paper, we theoretically study the free-energy landscape between the CB and W states of a single pore.  In particular, we assume that the liquid volume above the substrate is large enough to model the CB to W wetting transition from the intrusion and extrusion of the liquid into a single pore.  Therefore, our model can be used to study the stability of underwater superhydrophobicity of the substrate~\cite{Marmur2006,Xue2016,Xiang2017,Jones2017}.  We adopted the classical capillary model for truncated and inverted truncated conical pores, which are typical geometries to extract the effect of pore shape~\cite{Whyman2011,Checco2014,Jones2017,Giacomello2019}.  A detailed atomistic process,  which is beyond the scope of our classical capillary model, will be studied only numerically using the atomic simulations~\cite{Amabili2016,Giacomello2012,Savoy2012,Tinti2017,Jones2017,Prakash2016}, or the microscopic denisty functional theory~\cite{Singh2015,Giacomello2019}.  To make our model as simple as possible, we assume a flat liquid-vapor interface~\cite{Nosonovsky2007,Bormashenko2013}.  The line tension~\cite{Guillemont2012,Bormashenko2013,Iwamatsu2016,Schimmele2007,Law2017} at the liquid-vapor-solid triple line is neglected.  These two effects will be discussed briefly at the end of the next section

\section{\label{sec:sec2}Classic Capillary Model of Pore Wetting}

\subsection{Classical capillary model}

We consider the simplest wetting or drying transition by the intrusion and extrusion of liquid into a truncated conical pore~\cite{Jiang2020,Bormashenko2013,Jones2017,Checco2014}, as shown in Fig. 1 (a), and an inverted truncated conical pore~\cite{Jiang2020,Cao2007,Giacomello2019} with an overhanging structure, as shown in Fig. 1(b).  We do not consider a more complex scenario of wetting and drying in which a liquid droplet or a vapor bubble nucleates independently at the corner of the pore wall~\cite{Giacomello2012,Lv2015,Amabili2016,Jones2017}.  The nucleation of a liquid droplet or a vapor bubble is possible when the vapor pressure is high, or the liquid is highly stretched and metastable under negative pressure~\cite{Debenedetti1996}.

\begin{figure}[htbp]
%Fig.1
\begin{center}
\includegraphics[width=0.80\linewidth]{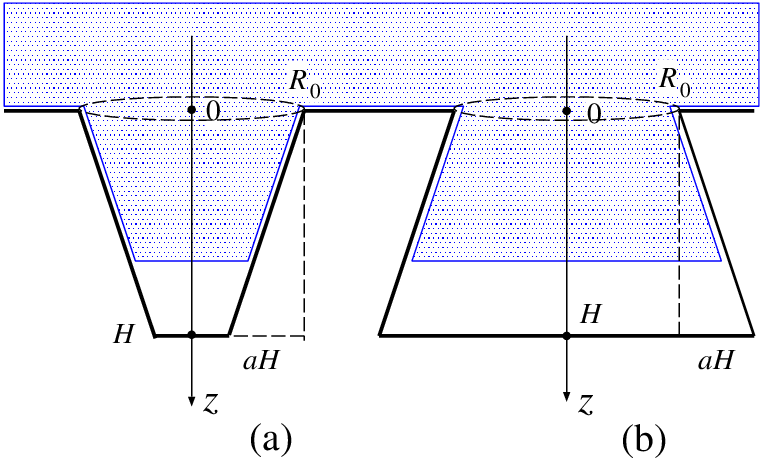}
\caption{
Two rotationally symmetric conical pores of  (a) a truncated cone with a narrowing radius and (b) an inverted truncated cone with a widening radius.  The opening radius and the depth of the conical pores are $R_{0}$ and $H$, respectively.  The rotational axis is $z$axis, and the inner wall has a constant slope $a$ that is represented as $f(z)=R_{0}\pm az$.
 }
\label{fig:1}
\end{center}
\end{figure}

Free-energy landscape is obtained from the grand potential $G$, which describes the thermodynamics of the liquid-vapor-solid (pore wall) system and depends on the chemical potential $\mu$ and on the temperature $T$~\cite{Amabili2016,Giacomello2012}, as:
\begin{equation}
G = F-\Delta p V
\label{eq:x1}
\end{equation}
by assuming a plausible transition pathway, where $V$ is the liquid volume inside the pore and $\Delta p$ is the pressure difference between the liquid and the vapor phase.  In fact, the vapor pressure can be the air pressure by adding the partial pressure of various gases in air to the vapor pressure of liquid.  Those gases trapped under the liquid will play an important role in the timescale of the CB to W transition which will be controlled by the diffusion of those gases.  However, we will not consider the effect of gases or air explicitly since the free energy landscape remain the same as long as the total pressure difference $\Delta p$ remains the same~\cite{Lv2015,Amabili2015}.

The first term in Eq.~(\ref{eq:x1}) given by
\begin{equation}
F=\gamma_{\rm lv}S_{\rm lv}-\gamma_{\rm lv}\cos\theta_{\rm Y}S_{\rm sl}
\label{eq:x2}
\end{equation}
is the surface free energy, where $\gamma_{\rm lv}$ and $S_{\rm lv}$ are the liquid-vapor surface tension,  surface area, respectively, and $S_{\rm sl}$ is the solid-liquid (wet) surface area.  The angle $\theta_{\rm Y}$ is the intrinsic Young's contact angle defined by the Young's equation
\begin{equation}
\cos\theta_{\rm Y}=\frac{\gamma_{sv}-\gamma_{\rm sl}}{\gamma_{\rm lv}}
\label{eq:x3}
\end{equation}
where $\gamma_{\rm sv}$ and $\gamma_{\rm sl}$ are the solid-vapor and solid-liquid surface tensions, respectively.

Bulk pressure difference $\Delta p$ between the liquid and vapor phases measures the relative stability of these two phases.  A positive pressure difference $\Delta p>0$ corresponds to the pressurized compressed liquid, which is more stable than the metastable vapor, while a negative pressure difference $\Delta p<0$ corresponds to the stretched metastable liquid and the vapor is stable~\cite{Debenedetti1996,Kelton2010}. A typical example of the former is the CB to W wetting transition on a underwater SHP substrate~\cite{Bormashenko2015,Xue2016,Jiang2020}, and the latter is bubble nucleation, called cavitation, in stretched liquids~\cite{Debenedetti1996,Kelton2010}.

We consider a pore with a rotationally symmetric shape around the $z$ axis, whose surface profile is given by
\begin{equation}
f(z)=R_{0} + d(z),
\label{eq:x4}
\end{equation}
where $R_{0}$ is the opening radius of the pore and $d(z)$ with $d(0)=0$ represents the surface topology of the pore wall (Fig. 1).

In this work, we consider the simplest topology of a straight inclined wall with $d(z)=\pm az$ which gives a truncated cone with a narrowing radius (Fig. 1(a)) and an inverted truncated cone with a widening radius (Fig. 1(b)) characterized by
\begin{equation}
f(z) = R_{0}\pm az,
\label{eq:x5}
\end{equation}
where $a>0$ is the slope of the wall.  The negative sign $-$ corresponds to the narrowing pore shown in Fig. 1(a) and the positive sign $+$ corresponds to the widening pore shown in Fig. 1(b).  The slope of a simple cylinder with a vertical wall is zero, i.e., $a=0$.

The truncated cone becomes a simple cone~\cite{Jones2017, Hauge1992, Reijmer1999,Malijevsky2015} when $a=R_{0}/H$.  This conical pore has been attracted special attentions~\cite{Hauge1992, Reijmer1999, Malijevsky2015} since the wetting transition called the filling transition can occur.  Due to the special geometry of the two-dimensional wedge and the three-dimensional cone, the cost of the surface free energy to move a flat liquid-vapor interface vanishes.  Then, the interface will be delocalized and move to infinity if the liquid and the vapor phases are thermodynamic equilibrium.  The thickness of the liquid layer adsorbed at the bottom of the wedge or cone will grow to infinity, and the complete wetting transition is realized.  Since the contact angle between the flat interface and the pore wall remains finite, this complete wetting transition with non-zero contact angle is called filling transition~\cite{Hauge1992, Reijmer1999, Malijevsky2015}.

In fact, the same filling transition can occur in our truncated cone when the liquid-vapor interface becomes flat.  Therefore, the filling transition from the bottom of truncated cone can occur at exactly the same contact angle as in the simple cone.  Similarly the intrusion of liquid from the top of turncated cone occurs without the cost of surface free energy.  However, those transiton occurs at the liquid-vapor thermodynamic  equilibrium.  Although, those wetting transition can also destroy CB state~\cite{Amabili2016,Jones2017,Giacomello2012}, we will concentrate on the simple intrusion and extrusion of liquid under the thermodynamic non-equilibrium condition $\Delta p \neq 0$ by the compression or the depression of liquid.

We consider the free-energy landscape of the liquid intrusion in (or extrusion from) the pore by assuming the following.  (1) The pinning of the liquid meniscus at the edge of the pore opening is not considered~\cite{Patanker2010,Papadopoulos2013}, (2) the liquid does not condense separately at the pore bottom~\cite{Giacomello2012,Jones2017,Kim2018}, and (3) the liquid-vapor meniscus is flat and horizontal~\cite{Nosonovsky2007,Bormashenko2013}.   Corrections due to non-flat meniscus and line tension at the triple line will be qualitatively considered later.

Our model, however, is too simplified for hydrophilic substrates because the pore wall is always wet and covered by a thin liquid layer~\cite{Reijmer1999}.  Also, the assumption of a flat meniscus is less accurate as the contact angle is low.  The interaction between the intruding liquid with the wetting layer enhanced by the curved meniscus will make our model less reliable, in particular, in the late stage of wetting when the pore is almost filled by the liquid.

We first consider the narrowing pore when the position of the liquid-vapor meniscus in the pore is $z$ (see Fig.~\ref{fig:1}).  The liquid volume $V$ in Eq.~(\ref{eq:x1}) is given by
\begin{equation}
V(z) = \pi \int_{0}^{z} f(z')^{2} dz'=\pi \left( R_{0}^{2}z- R_{0} a z^{2} + \frac{1}{3}a^{2}z^{3} \right),
\label{eq:x6}
\end{equation}
and the solid-liquid and the liquid-vapor surface areas are given by
\begin{eqnarray}
S_{\rm sl} &=& 2\pi \int_{0}^{z} f(z')\sqrt{1+\left(\frac{df}{dz'}\right)^{2}}dz'
\nonumber \\
&=& 2\pi R_{0}\sqrt{1+a^{2}}\left(z-\frac{1}{2}\frac{a}{R_{0}}z^{2}\right),
\label{eq:x7}
\end{eqnarray}
and
\begin{equation}
S_{\rm lv} = \pi f(z)^2=\pi\left(R_{0}-az\right)^{2},
\label{eq:x8}
\end{equation}
which gives the grand potential
\begin{equation}
G=\gamma_{\rm lv}\pi R_{0}^{2}g(\tilde{z})
\label{eq:x9}
\end{equation}
from Eq.~(\ref{eq:x1}) with the scaled free energy
\begin{eqnarray}
g(\tilde{z})&=&\left[\left(1-\alpha\tilde{z}\right)^2-2\cos\theta_{\rm Y}\sqrt{\eta^{2}+\alpha^{2}}\left(\tilde{z}-\frac{1}{2}\alpha\tilde{z}^{2}\right)\right]
\nonumber \\
&-&\tilde{p}\left[\tilde{z}-\alpha\tilde{z}^{2}+\frac{1}{3}\alpha^{2}\tilde{z}^{3}\right],
\label{eq:x10}
\end{eqnarray}
where
\begin{equation}
\tilde{z} = \frac{z}{H},\;\;\;(0\leq \tilde{z}\leq 1)
\label{eq:x11}
\end{equation}
and
\begin{equation}
\tilde{p}=\frac{H\Delta p}{\gamma_{\rm lv}},
\;\;\;\alpha=\frac{aH}{R_{0}},\;\;\;\eta=\frac{H}{R_{0}}.
\label{eq:x12}
\end{equation}
Therefore, the free-energy landscape is characterized by the scaled pressure $\tilde{p}$, and the shape of the pore is characterized by two parameters $\alpha$ and $\eta$, which represents the steepness of the wall, and the depth of the pore relative to the size of the opening, respectively (see Fig. 1).

It is straightforward to stud;y the inverted truncated cone with a widening radius (Fig. 1(b)) characterized by
\begin{equation}
f\left(z\right)=R_{0}+az.
\label{eq:x13}
\end{equation}
from Eq.~(\ref{eq:x5}).  We can use the formulaes derived above for the inverted truncated cone with a widening radius by changing the sign of $\alpha$ to negative.  A positive $\alpha$ ($\alpha>0$) corresponds to the narrowing pore and a negative alpha ($\alpha<0$) to the widening pore.

As the reference state, we consider the CB state when $z=0$, whose free-energy $g_{\rm CB}$ is
\begin{equation}
g_{\rm CB}=g(0)=1,\;\;\;{\rm or}\;\;\;G_{\rm CB}=\pi \gamma_{\rm lv}R_{0}^{2},
\label{eq:x14}
\end{equation}
and the free-energy difference $\Delta G(z)$ is
\begin{equation}
\Delta G(z)=G(z)-G_{\rm CB}=\gamma_{\rm lv}\pi R_{0}^{2}\Delta g(\tilde{z}).
\label{eq:x15}
\end{equation}
When $\Delta g>0$ the CB state is the most stable state and when $\Delta g<0$ the CB state would be metastable.  The scaled free energy $\Delta g(\tilde{z})$ from Eq.~(\ref{eq:x10}) becomes a cubic polynomial of $\tilde{z}$
\begin{equation}
\Delta g(\tilde{z}) = g_{1}\tilde{z} + g_{2}\tilde{z}^{2} + g_{3}\tilde{z}^{3},
\label{eq:x16}
\end{equation}
where
\begin{eqnarray}
g_{1} &=& -\left(2\alpha + 2\cos\theta_{\rm Y}\sqrt{\eta^{2}+\alpha^{2}} + \tilde{p} \right),
\label{eq:x17} \\
g_{2} &=& \alpha\left(\alpha + \cos\theta_{\rm Y}\sqrt{\eta^{2}+\alpha^{2}} + \tilde{p}\right),
\label{eq:x18} \\
g_{3} &=& -\frac{1}{3}\alpha^{2}\tilde{p}.
\label{eq:x19}
\end{eqnarray}

Although, the free energy described in Eq.~(\ref{eq:x9}) is correct as long as the liquid-vapor interface does not touch the pore bottom at $\tilde{z}=1$.  Once the liquid-vapor meniscus reaches the pore bottom, the free energy of the filled state given in Eqs.~(\ref{eq:x9}) and (\ref{eq:x10}) does not represent the free energy of the correct W state since the wetting of the pore bottom is not considered.  Therefore, our model assumes that the bottom surface is covered with a microscopic vapor layer, which might be true only when the bottom surface is strongly hydrophobic.  We call the state characterized by the free energy in Eqs.~(\ref{eq:x9})-(\ref{eq:x19}) with $\tilde{z}=1$ as the filled (F) state to distinguish it from the W state.  To consider the W state, we add the following correction  
\begin{equation}
\delta G=\left(\gamma_{\rm ls}-\gamma{\rm sv}-\gamma{\rm lv}\right)S_{\rm lv}\left(H\right)
\label{eq:x20}
\end{equation}
to the free-energies in Eqs.~(\ref{eq:x9}) and (\ref{eq:x10}), where $S_{\rm lv}\left(H\right)$ is the liquid-vapor surface area at the bottom given by
\begin{equation}
S_{\rm lv}\left(H\right)=\pi\left(R_{0}-aH\right)^{2}.
\label{eq:x21}
\end{equation}
from Eq.~(\ref{eq:x8}).  Therefore, we have to add the following correction
\begin{equation}
\delta g_{\rm W}=-\left(\cos\theta_{\rm Y}+1\right)\left(1-\alpha\right)^{2}
\label{eq:x22}
\end{equation}
to the scaled free energy $\Delta g_{\rm F}=g\left(\tilde{z}=1\right)$ of the F state to obtain the scaled free energy $\Delta g_{\rm W}$ of W state:
\begin{equation}
\Delta g_{\rm W}=\Delta g_{\rm F}+\delta g_{\rm W}.
\label{eq:x23}
\end{equation}
The correction $\delta g_{\rm W}$ can be interpreted as the adsorption energy of liquid at the bottom wall.  Apparently the adsorption energy $\delta g_{\rm W}$ is always negative since the liquid-vapor interface disappears. This adsorption energy $\delta g_{\rm W}$ vanishes only when the substrate is SHP with $\theta_{\rm Y}=180^{\circ}$.  The F state always has higher free energy than the W state since $\delta g_{\rm W}<0$.  Since the F state always acts as the free-energy barrier for the W state, the CB to W and the F to W transitions are always irreversible in our capillary model.

Therefore, the transition to the W state cannot be described properly in our model because the free energy is singular and jumps at $\tilde{z}=1$~\cite{Patanker2004,Checco2014}. A more microscopic description~\cite{Checco2014,Jones2017,Giacomello2012,Kim2018} than the macroscopic capillary model to trace the process of wetting of the bottom wall is necessary to describe the W state.  

It is also possible, to study the final stage of wetting or the initial state of dewetting of the pore bottom using the classical capillary model by assuming the liquid or vapor nucleation, for example, from the bottom corner of the pore~\cite{Lv2015,Giacomello2019,Kim2018}.  In particular, the macroscopic theory of the filling transition~\cite{Hauge1992, Reijmer1999, Malijevsky2015} can be applicable when the pore is rectangular and the bottom corner is the wedge formed by three orthogonal planes.  In our work, however,  We only focus on the global free-energy landscape between the CB and F states.

\subsection{Critical pressure and critical Young's contact angle}

The stability limit of the SHP-CB state is determined from $d\Delta g /d\tilde{z}=0$ or $g_{1}=0$ at $\tilde{z}=0$, which gives the critical pressure $\tilde{p}_{c}$:
\begin{equation}
\tilde{p}_{c} = -\left(2\alpha + 2\cos\theta_{\rm Y}\sqrt{\eta^{2}+\alpha^{2}} \right),
\label{eq:x24}
\end{equation}
which is written as
\begin{equation}
\Delta p_{\rm c}=-\frac{2\gamma_{\rm lv}\left(a+\cos\theta_{\rm Y}\sqrt{1+a^{2}}\right)}{R_{0}}
\label{eq:x25}
\end{equation}
using the original unit from Eq.~(\ref{eq:x12}).  When the pore is a cylinder with straight vertical wall, we obtain the well known formula of the force balance of the Laplace pressure
\begin{equation}
\Delta p_{c}=-\frac{2\gamma_{\rm lv}\cos\theta_{\rm Y}}{R_{0}}
\label{eq:x26}
\end{equation}
by setting $a=0$ in Eq.~(\ref{eq:x25}).

\begin{figure}[htbp]
\begin{center}
%Fig.2
\subfigure[]
{
\includegraphics[width=0.8\linewidth]{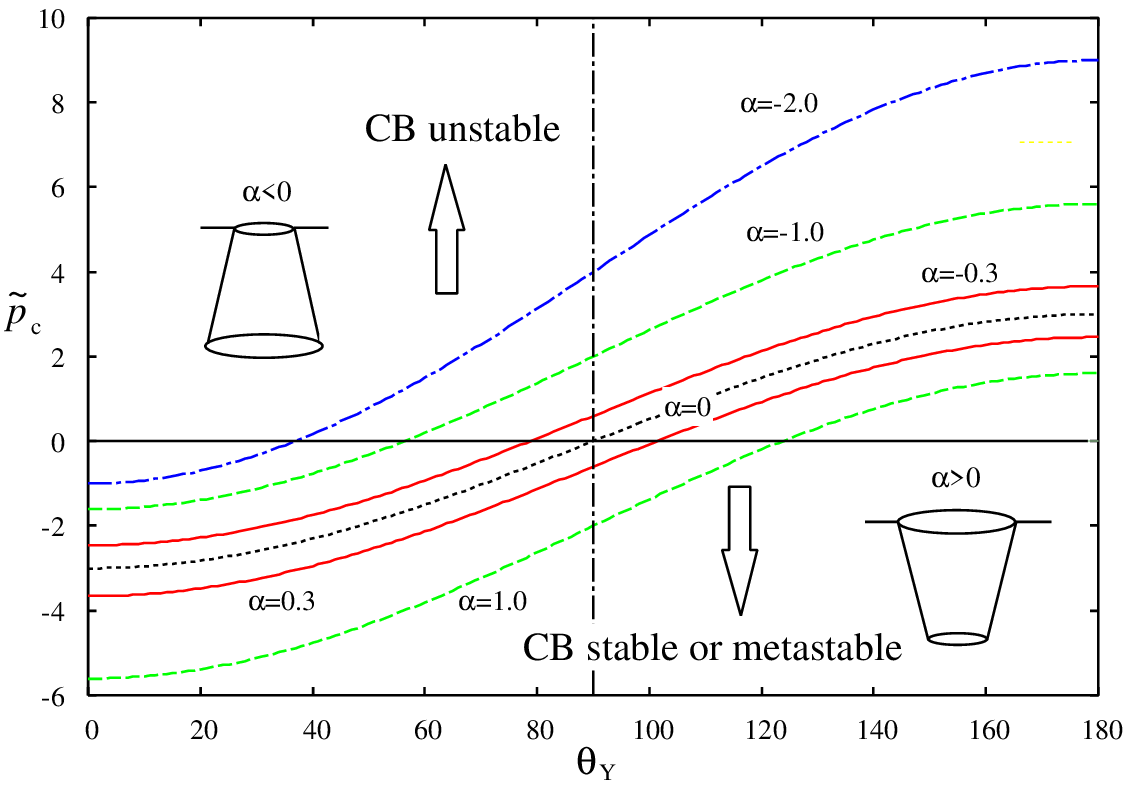}
\label{fig:2a}
}
\subfigure[]
{
\includegraphics[width=0.8\linewidth]{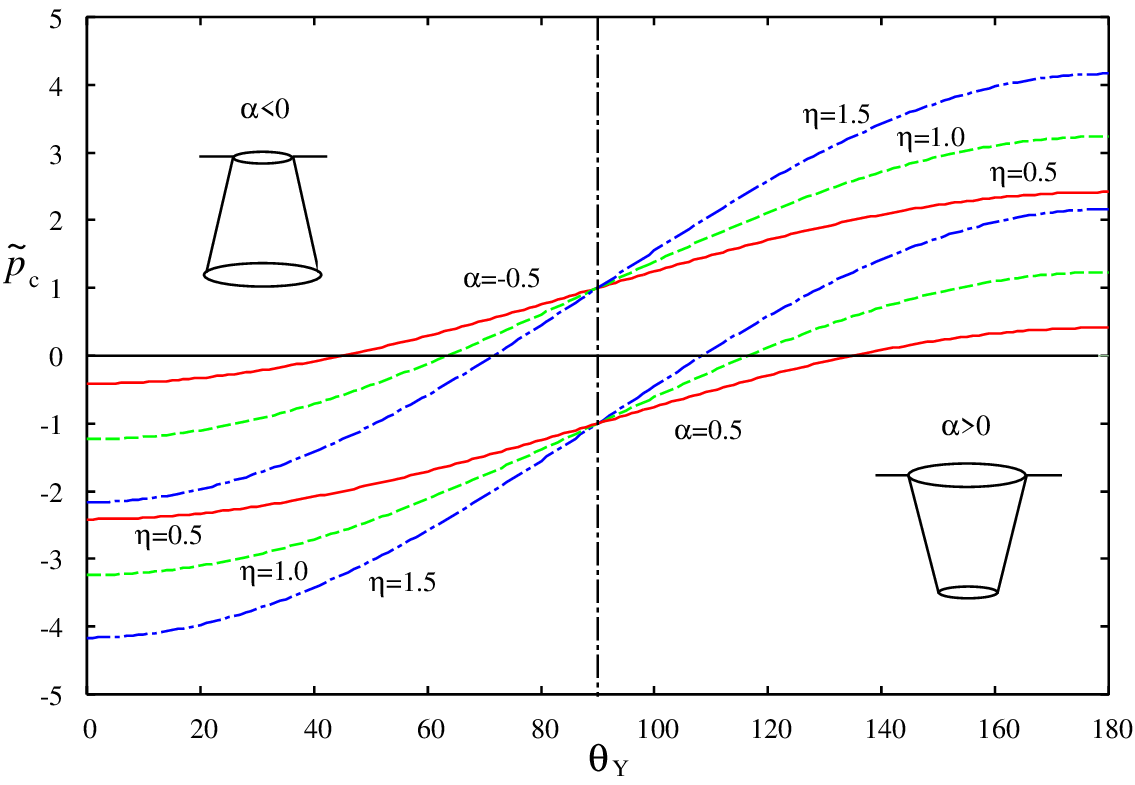}
\label{fig:2b}
}
\end{center}
\caption{
The critical pressure $\tilde{p}_{\rm c}$ as a function of Young's contact angle $\theta_{\rm Y}$. The CB state will be unstable above these curves when $\tilde{p}>\tilde{p}_{\rm c}$.  (a) For various slope factors $\alpha$,  the bottom two curves with $\alpha>0$ correspond to the narrowing pores, the top three curves with $\alpha<0$ correspond to the widening pores, and $\alpha=0$ corresponds to a straight cylindrical pore.  (b) For various depths $\eta$, the bottom three curves correspond to the narrowing pores, and the top three curves correspond to the widening pores.  For a large $\theta_{\rm Y}$, the critical pressure will be positive and large.  It becomes positive even when $\theta_{\rm Y}<90^{\circ}$ for the widening pores with $\alpha<0$, which means that the SHP-CB state will be possible under compressed liquid even when the substrate is hydrophilic.  The critical Young's contact angle $\theta_{\rm c}$ in Fig.~\ref{fig:3} is determined from the zero of $\tilde{p}_{\rm c}=0$. 
} 
\label{fig:2}
\end{figure}

\begin{figure}[htbp]
\begin{center}
%Fig.3
\subfigure[]
{
\includegraphics[width=0.8\linewidth]{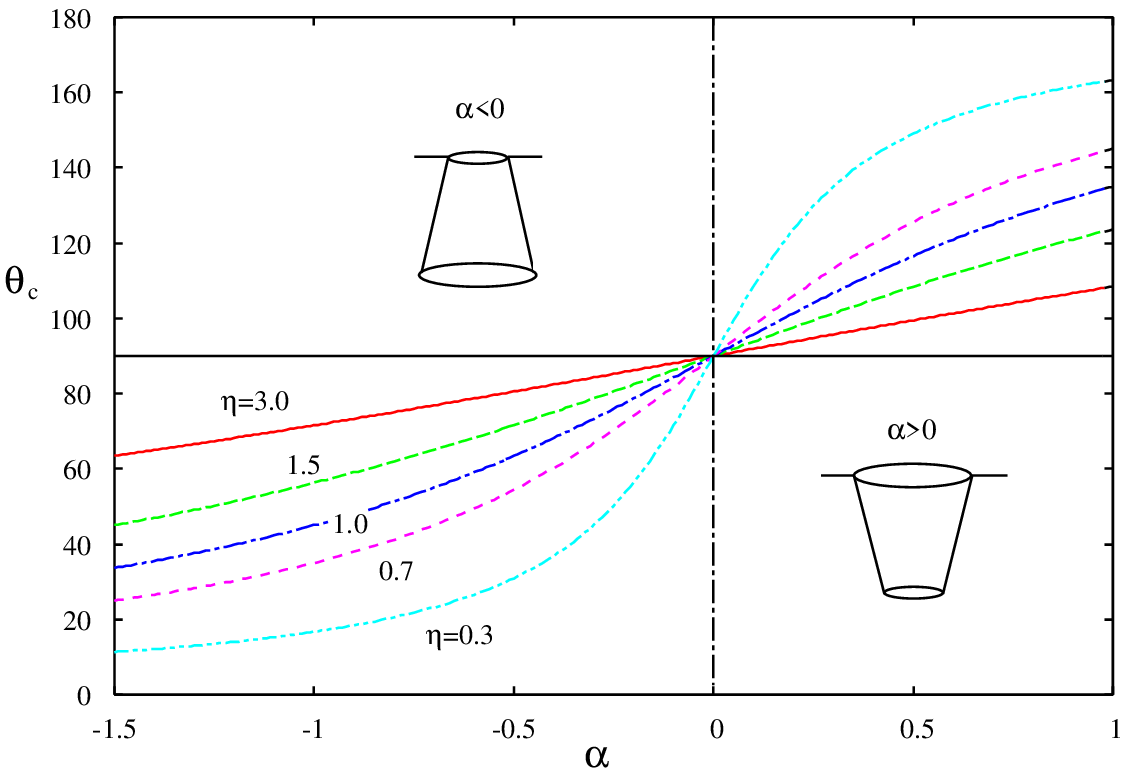}
\label{fig:3a}
}
\subfigure[]
{
\includegraphics[width=0.8\linewidth]{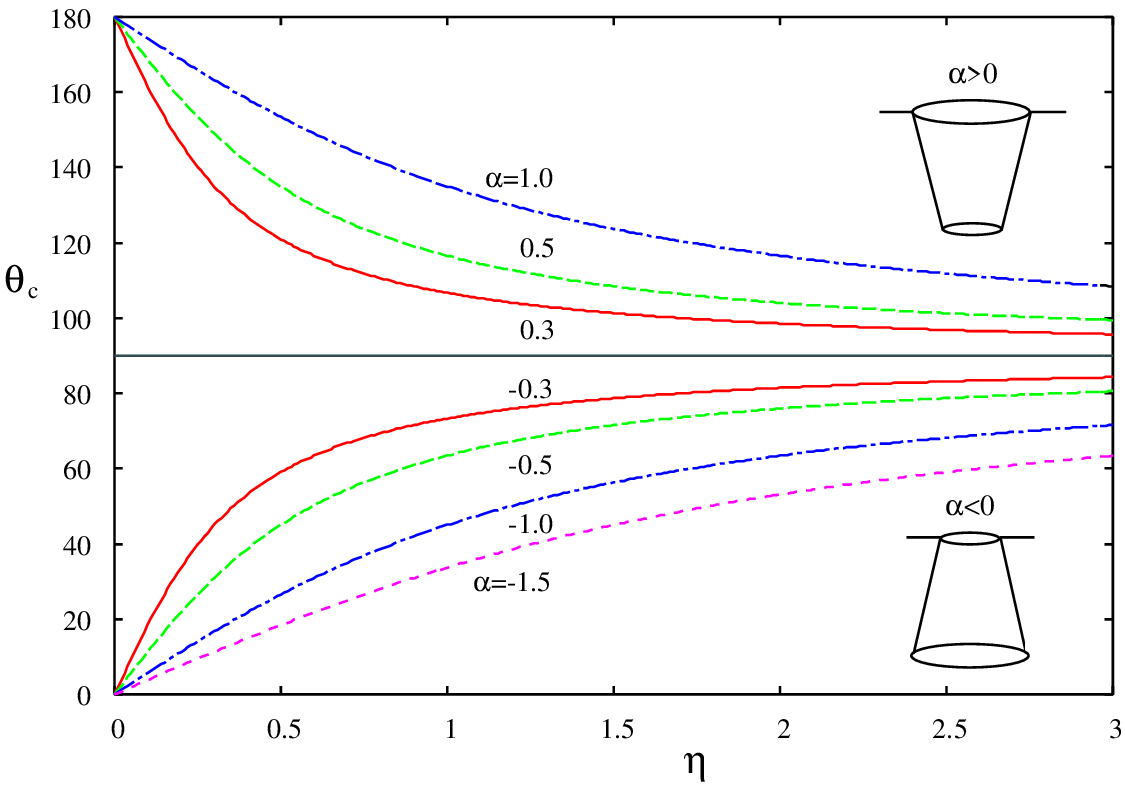}
\label{fig:3b}
}
\end{center}
\caption{
The critical Young's contact angle $\theta_{\rm c}$ defined by $\tilde{p}_{\rm c}=0$ as a function of (a) the shape factor $\alpha$ and (b) the pore depth $\eta$.  The CB state would be stable under the hydrostatic pressure  with $\tilde{p}<\tilde{p}_{\rm c}$ (see Fig.~\ref{fig:2a}) when $\theta_{\rm Y}>\theta_{\rm c}$.  The critical angle $\theta_{\rm c}$ is always smaller than $90^{\circ}$ when $\alpha<0$ (widening pore), while it is always larger than $90^{\circ}$ when $\alpha>0$ (narrowing pore).  It approaches the neutral contact angle $90^{\circ}$ when the pore depth or $\eta$ approaches infinity.
 } 
\label{fig:3}
\end{figure}

Figure \ref{fig:2a} shows the critical pressure $\tilde{p}_{\rm c}$ as a function of Young's contact angle $\theta_{\rm Y}$ for various shape factors $\alpha$ when $\eta=1.5$ ($H=1.5 R_{0}$).  The SHP-CB state becomes unstable when $\tilde{p}>\tilde{p}_{\rm c}$.  The submerged pore under the compressed pressurized liquid corresponds to $\tilde{p}_{\rm c}>0$, while that under the decompressed metastable stretched liquid corresponds to $\tilde{p}_{\rm c}<0$.  The critical pressure becomes positive, i.e., $\tilde{p}_{\rm c}>0$ only when $\theta_{\rm Y}>\theta_{\rm c}$ (critical Young's contact angle), where
\begin{equation}
\theta_{\rm c}=\cos^{-1}\frac{-\alpha}{\sqrt{\eta^{2}+\alpha^{2}}}
\label{eq:x27}
\end{equation}
is determined from $\tilde{p}_{\rm c}\left(\theta_{\rm Y}\right)=0$ in Eq.~(\ref{eq:x24}).  This critical angle is the lower-bound of $\theta_{\rm Y}$ for which the CB state is stable under the compressed liquid since the CB state is stable as long as $\tilde{p}_{\rm c}>\tilde{p}>0$ (Fig.~\ref{fig:2a}).  This critical angle is alway larger than $90^{\circ}$ in the narrowing pore with $\alpha>0$, while it is always smaller than $90^{\circ}$ in the widening pore with $\alpha<0$ in Fig.~\ref{fig:2a} (see also Fig.~\ref{fig:3a} ).

Therefore, a hydrophobic surface with $\theta_{\rm Y}>\theta_{\rm c}>90^{\circ}$ is always necessary to achieve the SHP-CB state under the compressed liquid in the narrowing pore, while even a hydrophilic surface with $90^{\circ}>\theta_{\rm Y}>\theta_{\rm c}$ is possible in the widening pore~\cite{Xue2016,Marmur2006}.   Further, the larger the magnitude of the slope $|\alpha|$ the higher the critical pressure $\tilde{p}_{\rm c}$ in the widening pore.  Therefore, an ink-bottle shape with a narrow neck and a wide bottom would be favorable for the stability of the SHP-CB state~\cite{Xue2016,Jiang2020,Marmur2006}.

Figure \ref{fig:2b} shows the critical pressure $\tilde{p}_{\rm c}$ as a function of $\theta_{\rm Y}$ for various pore depths $\eta$ when $\alpha=0.5$ and $\alpha=-0.5$.   The critical pressure is more sensitive to $\theta_{\rm Y}$ when the parameter $\eta$ is large or the pore is deep.  Apparently, a deep ($\eta=1.5$) widening pore ($\alpha=-0.5$) of hydrophobic substrate ($\theta_{\rm Y}>90^{\circ}$) is the most favorable to increase the critical pressure $\tilde{p}_{\rm c}$.

Figure \ref{fig:3a} shows the critical contact angle $\theta_{\rm c}$ as a function of the shape factor $\alpha$ for varying depths $\eta$.  The critical pressure becomes positive $\tilde{p}_{\rm c}>0$ when $\theta_{\rm Y}>\theta_{\rm c}$.  The critical contact angle $\theta_{\rm c}$ is always larger than $90^{\circ}$ when $\alpha>0$, while it can be smaller than $90^{\circ}$ when $\alpha<0$.  Therefore an inverted truncated cone  ($\alpha<0$) can sustain the CB state even when the substrate is hydrophilic with $\theta_{\rm c}<\theta_{\rm Y}<90^{\circ}$.  The critical angle is less sensitive to $\alpha$ when the pore is deep (e.g., $\eta=3.0$) while it is more sensitive to $\alpha$ when the pore is shallow (e.g., $\eta=0.3$).

Figure \ref{fig:3b} shows the critical contact angle $\theta_{\rm c}$ as a function of the pore depth of $\eta$.  Apparently $\theta_{\rm c}>90^{\circ}$ for $\alpha>0$ and $\theta_{\rm c}<90^{\circ}$ for $\alpha<0$.  A shallow (small $\eta$) inverted truncated pore ($\alpha<0$) with a large $\left|\alpha\right|$ is favorable to reduce $\theta_{\rm c}$ and stabilize the CB state under compressed liquid even when the substrate is hydrophilic.

\subsection{Free energy landscape}

The critical pressure $\tilde{p}_{\rm c}$ and, the critical contact angle $\theta_{\rm c}$ only indicate the stability of the CB state.  It does not necessarily mean that when $\tilde{p}>\tilde{p}_{\rm c}$, the CB state is absolutely unstable and the completely wet W state is absolutely stable.  For instance, it could be possible to obtain an incompletely filled pore.  Moreover, even when the CB state loses its stability, it can remain metastable through a free-energy barrier.  Therefore, the information on the free-energy landscape and the free-energy barrier between the CB and the W states would be necessary to evaluate the stability of the CB state.  This would be helpful in designing sustainable SHP.

\begin{figure}[htbp]
\begin{center}
%Fig.4
\subfigure[]
{
\includegraphics[width=0.8\linewidth]{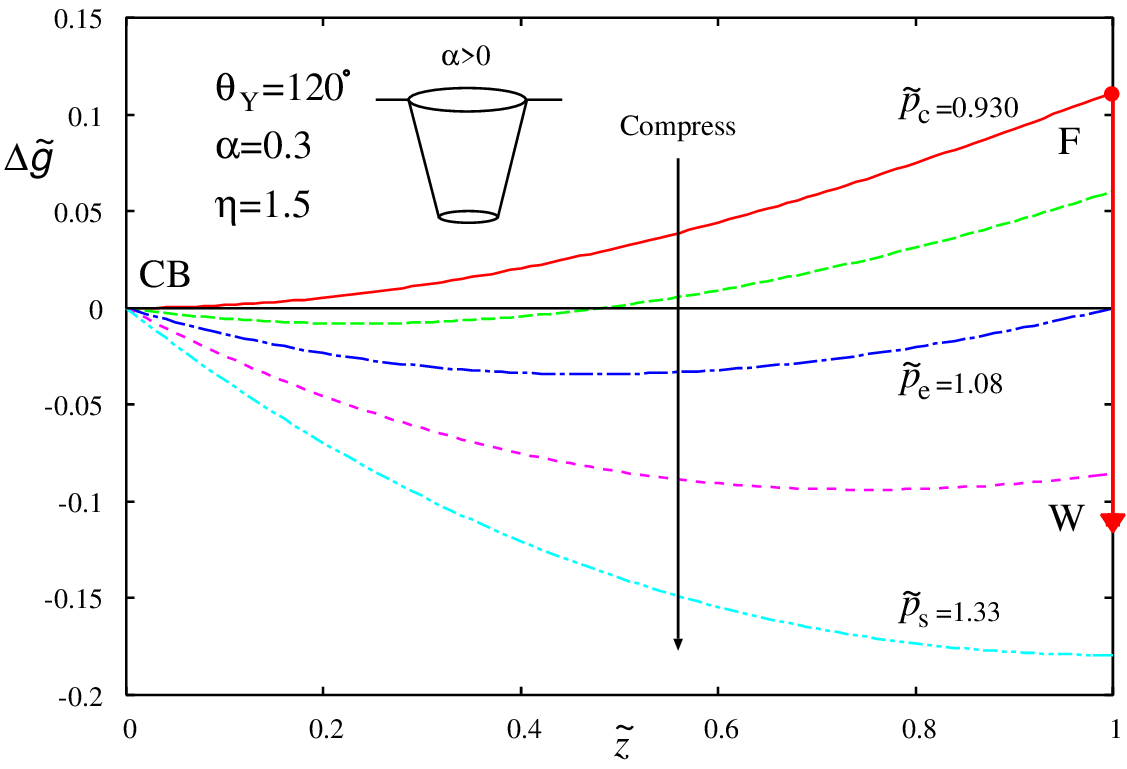}
\label{fig:4a}
}
\subfigure[]
{
\includegraphics[width=0.8\linewidth]{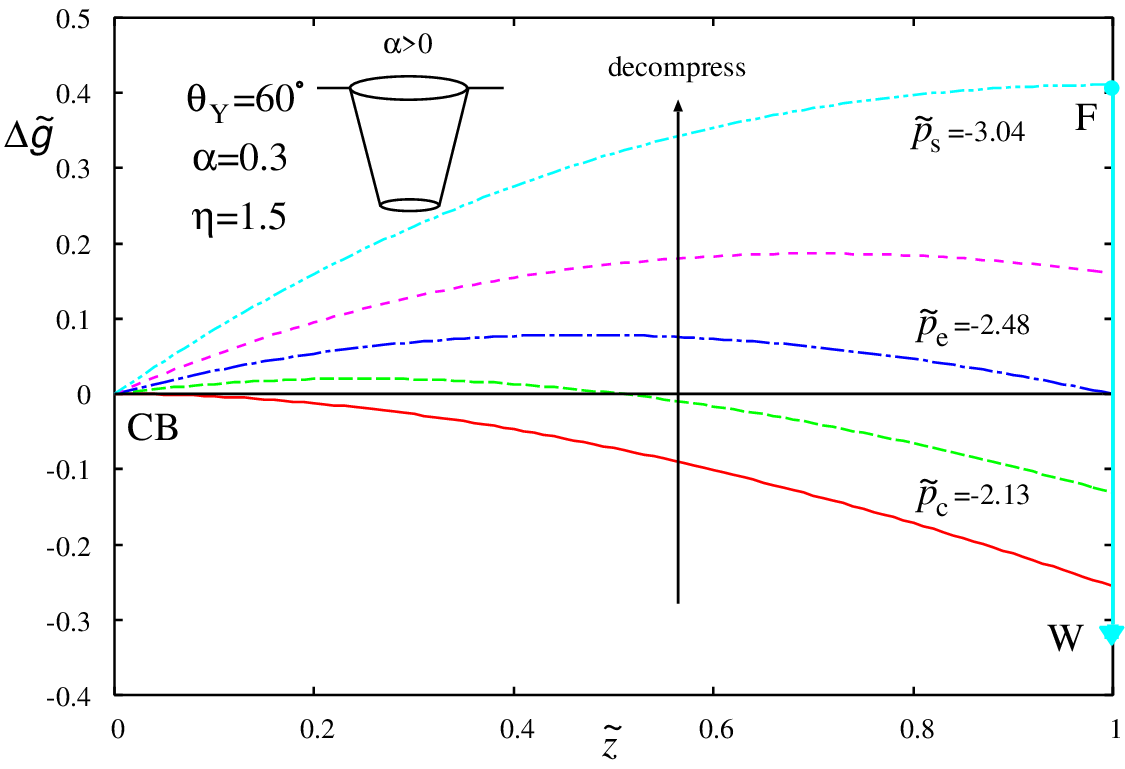}
\label{fig:4b}
}
\end{center}
\caption{
The free-energy landscape between $\tilde{z}=0$ (CB state) and $\tilde{z}=1$ (F state) of a truncated cone with $\alpha=0.3$ and $\eta=1.5$ when the substrate is (a) hydrophobic ($\theta_{\rm Y}=120^{\circ}>\theta_{\rm c}=101.3^{\circ}$) and (b) hydrophilic ($\theta_{\rm Y}=60^{\circ}<\theta_{\rm c}=101.3^{\circ}$).  The adsorption energy in (a) is $\delta g_{\rm W}=-0.245$ and in (b) is $\delta g_{\rm W}=-0.735$ indicated by the down arrow at $\tilde{z}=1$.  Therefore, the free energy of the W state $\Delta g_{\rm W}$ is always lower than that of the filled F state with $\Delta g_{\rm F}=\Delta g\left(\tilde{z}=1\right)$.  In (a) the applied pressure is $\tilde{p}=\tilde{p}_{\rm c} (0.930), 1.00, \tilde{p}_{\rm e} (1.08), 1.20, \tilde{p}_{\rm s} (1.33)$, and  in (b) the applied pressure is $\tilde{p}=\tilde{p}_{\rm s} (-3.04), -2.70, \tilde{p}_{\rm e} (-2.48), -2.30, \tilde{p}_{\rm c} (-2.13)$ from the top to the bottom. 
 } 
\label{fig:4}
\end{figure}

Since the free-energy difference in Eq. (\ref{eq:x16}) is a cubic polynomial of $\tilde{z}$ given by
\begin{equation}
\Delta g\left(\tilde{z}\right)=\left(\tilde{p}_{\rm c}-\tilde{p}\right)\tilde{z}
-\alpha\left(\frac{\tilde{p}_{\rm c}}{2}-\tilde{p}\right)\tilde{z}^{2}-\frac{1}{3}\alpha^{2}\tilde{p}\tilde{z}^{3},
\label{eq:x28}
\end{equation}
it has two extrema at $\tilde{z}_{1}$ and $\tilde{z}_{2}$ determined from $d\Delta g/d\tilde{z}=0$.  They are
\begin{equation}
\tilde{z}_{1}=\frac{1}{\alpha},
\label{eq:x29}
\end{equation}
which is unphysical because $z_{1}\ge 1$ when $1\ge\alpha\ge 0$ and $z_{1}\leq 0$ when $\alpha\leq 0$, and
\begin{eqnarray}
\tilde{z}_{2} &=& \tilde{z}_{\rm ex}=\frac{1}{\alpha}\left(1-\frac{\tilde{p}_{\rm c}}{\tilde{p}}\right),
\label{eq:x30}
\end{eqnarray}
which is a physically meaningful extremum, i.e.,  $\tilde{z}_{\rm ex}=\tilde{z}_{2}$ if $\tilde{p}>\tilde{p}_{\rm c}$ when $\alpha>0$ and $\tilde{p}<\tilde{p}_{\rm c}$ when $\alpha<0$.  Therefore, the free energy in Eq.~(\ref{eq:x28}) has only one extremum at $\tilde{z}_{\rm ex}$ where the free energy becomes
\begin{equation}
\Delta g_{\rm ex}=-\frac{\left(\tilde{p}-\tilde{p}_{\rm c}\right)^{2}\left(2\tilde{p}+\tilde{p}_{\rm c}\right)}{6\alpha\tilde{p}^{2}},
\label{eq:x31}
\end{equation}
which corresponds to the minimum when $\Delta g_{\rm ex}<0$ and the maximum when $\Delta g_{\rm ex}>0$.  Note that $\Delta g=0$ when $\tilde{z}=0$ (CB state).

At the extremum given by Eq.~(\ref{eq:x30}), the radius of the circular contact line $R_{\rm ex}=f\left(\tilde{z}_{\rm ex}\right)$ is given by
\begin{equation}
R_{\rm ex}=R_{0}\left(1-\alpha\tilde{z}_{\rm ex}\right)=R_{0}\frac{\tilde{p}_{\rm c}}{\tilde{p}},
\label{eq:x32}
\end{equation}
which gives an equation similar to that of the Laplace pressure
\begin{equation}
R_{\rm ex}=-\frac{2\gamma_{\rm lv}\left(a+\cos\theta_{\rm Y}\sqrt{1+a^{2}}\right)}{\Delta p}
\label{eq:x33}
\end{equation}
in the original unit in Eq.~(\ref{eq:x12}).

Figure~\ref{fig:4a} shows the free-energy landscape between $\tilde{z}=0$ (CB state) and $\tilde{z}=1$ (F state and not exactly the W state) of a truncated conical pore with a narrowing radius of $\alpha=0.3$ and $\eta=1.5$.  Further, the pore wall is made up of a hydrophobic non-wettable substrate with $\theta_{\rm Y}=120^{\circ}>\theta_{\rm c}=101.3^{\circ}$.  The free energy  $\Delta g_{\rm W}=\Delta g_{\rm F}+\delta g_{\rm W}$ of the W sate is always lower than that of the F state $\Delta g_{\rm F}=\Delta g(z=1)$ by the amount $\delta g_{\rm W}=-0.245$ from Eq.~(\ref{eq:x22}), which is large in the scale of Fig.~\ref{fig:4a}.

By compressing the liquid from zero pressure, the CB state becomes unstable at the critical pressure $\tilde{p}_{\rm c}=0.930$. Even at this pressure, the free energy of the F state is positive ($\Delta g_{\rm F}=\Delta g\left(\tilde{z}=1\right)=0.112$); however, the free energy of the W state becomes negative ($\Delta g_{\rm W}=\Delta g_{\rm F}+\delta g_{\rm W}=0.112-0.245=-0.113$).  Therefore, the W state has a lower free energy as compared to that of the CB state.  Therefore, the free-energy of the F state acts as the energy barrier to prevent the CB to W transition (Fig.\ref{fig:4a}).

As the pressure $\tilde{p}$ increase, the liquid starts to intrude into the pore but does not completely fill the pore. Instead, the liquid-vapor meniscus is trapped at the free-energy minimum (Fig.~\ref{fig:4a}) given by Eq.~(\ref{eq:x30}).  With the further increased in pressure, the liquid further intrude into the pore.   Then, the free energies at $\tilde{z}=0$ (CB state) and at $\tilde{z}=1$ (F state) become equal at $\tilde{p}_{\rm e}=1.08$ when $\Delta g_{\rm F}=\Delta g(\tilde{z}=1)=0$, which gives
\begin{equation}
\tilde{p}_{\rm e}=\frac{3\left(2-\alpha\right)}{2\left(3-3\alpha+\alpha^{2}\right)}\tilde{p}_{\rm c}.
\label{eq:x34}
\end{equation}
Equation (\ref{eq:x34}) does not represent the pressure when the free energies of the CB state and W states become equal.  This pressure $\tilde{p}_{\rm CB-W}$ is determined from $\Delta g_{\rm W}=0$ in Eq.~(\ref{eq:x23}), which gives
\begin{equation}
\tilde{p}_{\rm CB-W}=\tilde{p}_{e}-\frac{3\left(\cos\theta_{\rm Y}+1\right)\left(1-\alpha\right)^{2}}{3-3\alpha+\alpha^{2}}.
\label{eq:x35}
\end{equation}
Note that we always have $\tilde{p}_{\rm e}\leq\tilde{p}_{\rm CB-W}$.  The equality $\tilde{p}_{\rm e}=\tilde{p}_{\rm CB-W}$ holds only when $\theta_{\rm Y}=180^{\circ}$.  Figure~\ref{fig:4a} indicates that the W state cannot be attained even at $p_{\rm e}$ because the F state acts as the energy barrier.

\begin{figure}[htbp]
\begin{center}
%Fig.4
\subfigure[]
{
\includegraphics[width=0.8\linewidth]{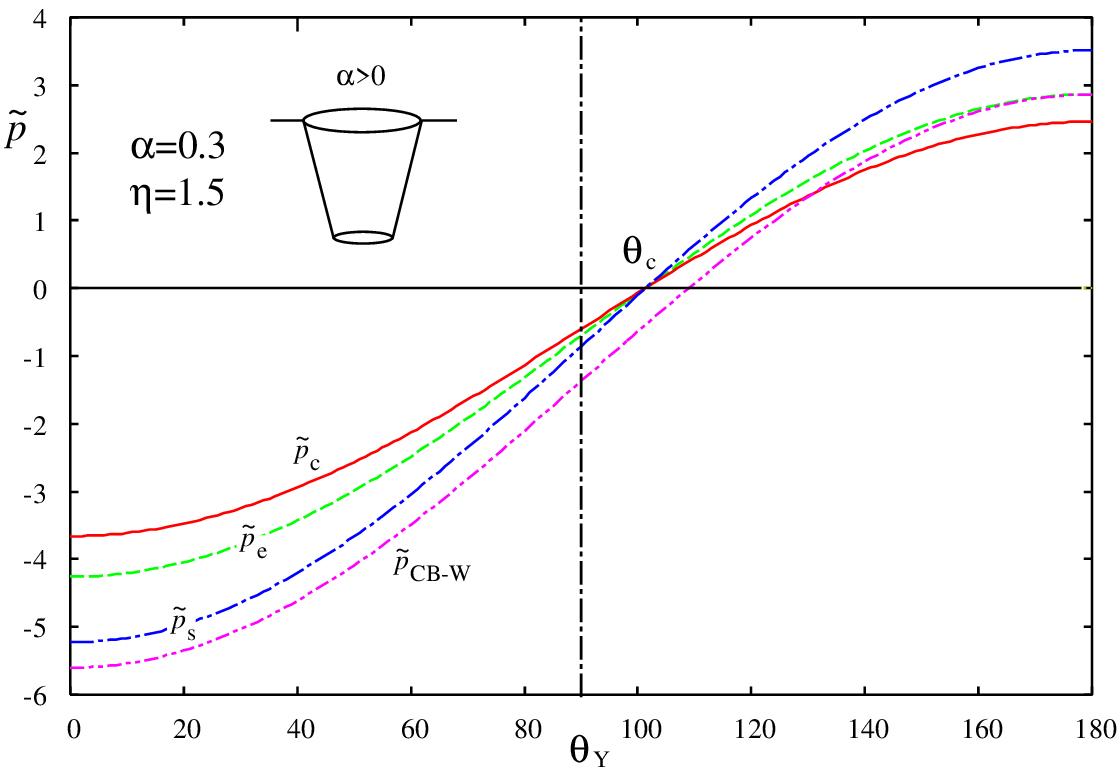}
\label{fig:5a}
}
\subfigure[]
{
\includegraphics[width=0.8\linewidth]{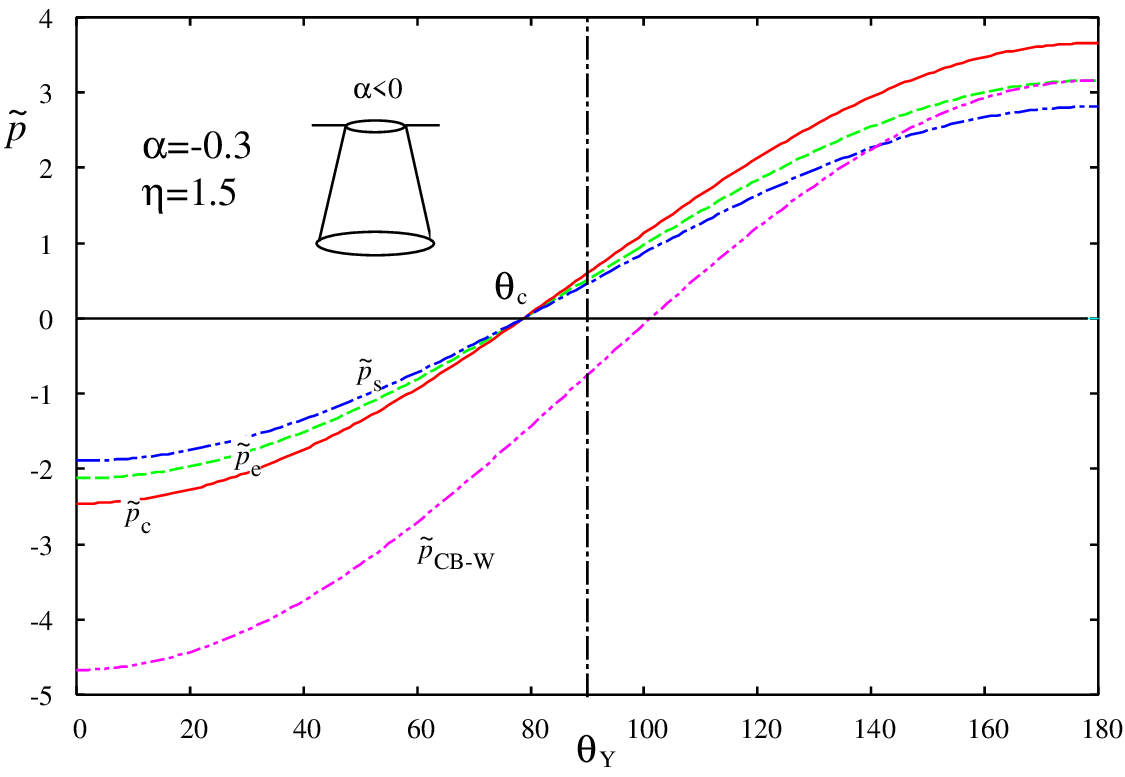}
\label{fig:5b}
}
\end{center}
\caption{
Three pressures $\tilde{p}_{\rm c}$, $\tilde{p}_{\rm e}$, and $\tilde{p}_{\rm s}$, which characterize the free-energy landscape, and $\tilde{p}_{\rm CB-W}$ for (a) the truncated cone with $\alpha=0.3$ and (b) the inverted truncated cone with $\alpha=-0.3$ (upper three curves) when $\eta=1.5$.  Three pressures are always positive when $\theta_{\rm Y}>\theta_{\rm c}$ and negative when $\theta_{\rm Y}<\theta_{\rm c}$. Their oder also changes at $\theta_{\rm c}$.  The different order of those three pressures leads to the completely different and inverted free-energy landscapes in Fig.~\ref{fig:4} and \ref{fig:6}.  The pressure $\tilde{p}_{\rm CB-W}$ is generally smaller than the three pressures.
 } 
\label{fig:5}
\end{figure}

Figures~\ref{fig:4a} and \ref{fig:5a} indicate that there is a possibility of the first order phase transition to the W state not from the completely dry CB state but from the partially filled CB state at the free-energy minimum $z_{e}$ given by Eq. ~(\ref{eq:x30}). The pressure $\tilde{p}$ where this transition become possible will be determined from $\Delta g_{W}=\Delta g_{\rm ex}$ from Eqs.~(\ref{eq:x23}) and (\ref{eq:x31}), which leads to the cubic equation for the pressure $\tilde{p}$ and the solution which satisfies $\tilde{p}_{c}<\tilde{p}<\tilde{p}_{\rm s}$ corresponds to the pressure at which the first order transition from the partially filled CB state to the completely filled W state take place.  We will not consider this problem further and will concentrate on the global character of the free-energy landscape.

The depth of the free-energy minimum near $\tilde{z}=0.5$ can be estimated from Eq.~(\ref{eq:x31}) which gives $\Delta g_{\rm ex}\simeq -0.034$ at $\tilde{p}_{\rm e}=1.08$ in the scaled unit and $\left|\Delta G_{\rm ex}\right|=0.034\pi \gamma_{\rm lv}R_{0}^{2}$ in the original unit (Eq.~(\ref{eq:x15})).  Suppose $R_{0}=10$ nm and $\gamma_{\rm lv}=73$mJm$^{-1}$ (water), we then obtain $\left|\Delta G_{\rm ex}\right|\simeq 7.8\times 10^{-19}$J, which is two orders of magnitude larger than the thermal energy $kT$.  As a result, the thermal fluctuation of the liquid-vapor interface would be small and the interface would be trapped at the free-energy minimum.

When the applied pressure $\tilde{p}$ increases beyond $\tilde{p}_{\rm e}=1.08$, the liquid-vapor interface moves toward $\tilde{z}=1$; however it is still trapped at the free-energy minimum given by Eq.~(\ref{eq:x30}).  Finally, when the position of the minimum reaches the pore bottom and $\tilde{z}_{\rm ex}=1$ in Eq.~(\ref{eq:x30}), which is realized by the pressure $\tilde{p}_{\rm s}$ given by
\begin{equation}
\tilde{p}_{\rm s}=\frac{\tilde{p}_{\rm c}}{1-\alpha},
\label{eq:x36}
\end{equation}
the CB state reaches the real stability limit; thus the F state with $\tilde{z}=1$ will be realized. Subsequently, the pore bottom would be wet, and the F to W transition would occur.  Consequently, the CB to W transition would occur because the W state always has a lower free energy that the F state ($\Delta g_{\rm W}<\Delta g_{\rm F}$).  Hence, the pore would be completely filled and wet.

The pressure $\tilde{p}_{\rm s}$ is higher for a larger $\alpha$ (Eq.~(\ref{eq:x36})) and becomes infinitely large for a conical-shaped pore~\cite{Jones2017} with $\alpha=1$.  The free-energy landscape in Fig.~\ref{fig:4a} provides a complete picture of the intrusion of the liquid into the truncated conical pore.  In other words, it describes the collapse of the SHP-CB state from the applied pressure.  The CB state does not collapse suddenly at $\tilde{p}_{\rm c}$, which is indicated by the classic formula in Eq.~(\ref{eq:x25}).  Rather, the collapse occurs by the gradual intrusion of the liquid-vapor meniscus.  The completely filled and wet W state might be realized at $\tilde{p}_{\rm s}>\tilde{p}_{\rm c}$ but not at $\tilde{p}_{\rm c}$.  This collapse of the CB state is irreversible because the adsorption energy $\delta g_{\rm W}$ acts as the energy barrier in the W to F dewetting transition.

Figure~\ref{fig:4b} shows the free-energy landscape between $\tilde{z}=0$ (the CB state) and $\tilde{z}=1$ (the F state) of a truncated conical pore with a narrowing radius $\alpha=0.3$ and made up of a hydrophilic wettable substrate with $\theta_{\rm Y}=60^{\circ}<\theta_{\rm c}=101.3^{\circ}$ and $\eta=1.5$.  The free energy of the W sate is always lower than that of the F state by the amount $\delta g_{\rm W}=-0.735$ obtained from Eq.~(\ref{eq:x22}), which has magnitude larger than that of the hydrophobic substrate in Fig.~\ref{fig:4a} (-0.245).

The convex free-energy landscape shown in Fig.~\ref{fig:4b} is completely different from the concave landscape in Fig.~\ref{fig:4a}.  When the liquid is weakly decompressed, the F state remains stable; however, the unstable CB state becomes metastable at $\tilde{p}_{\rm c}=-2.13$.  When the liquid is further decompressed, a free-energy barrier between the CB and F states starts to develop.  The free energies of the CB and F states become equal at $\tilde{p}_{\rm e}=-2.70$ (Fig.~\ref{fig:4b}).  However, there remains a free-energy barrier of height similar in magnitude to the depth of the free-energy well shown in Fig.~\ref{fig:4a}.  Since the free energy of the W state is always lower than that of the CB state, the W state is stabler than the CB state at $\tilde{p}_{\rm e}$.  As the pressure $\tilde{p}$ becomes more negative, the CB state becomes more stable compared to the F state; however, the W state remains the most stable.

Finally, the free-energy barrier between the CB and F states vanishes, and the F state has a higher free energy $\Delta g_{\rm F}=\Delta g\left(\tilde{z}=1\right)=0.411$ than that of the CB state at $\tilde{p}_{\rm s}=-3.04$, which is similar to the spinodal of the first-order phase transition~\cite{Debenedetti1996,Kelton2010}.  However, the free energy of the W state is $\Delta g_{\rm W}=0.411-0.735=-0.324$; hence,  the W sate is still more stable than the CB state.  Therefore, the spontaneous transition from the W state to the CB state will not occur, and the SHP-CB state will not be recovered.

\begin{figure}[htbp]
\begin{center}
%Fig.6
\subfigure[]
{
\includegraphics[width=0.8\linewidth]{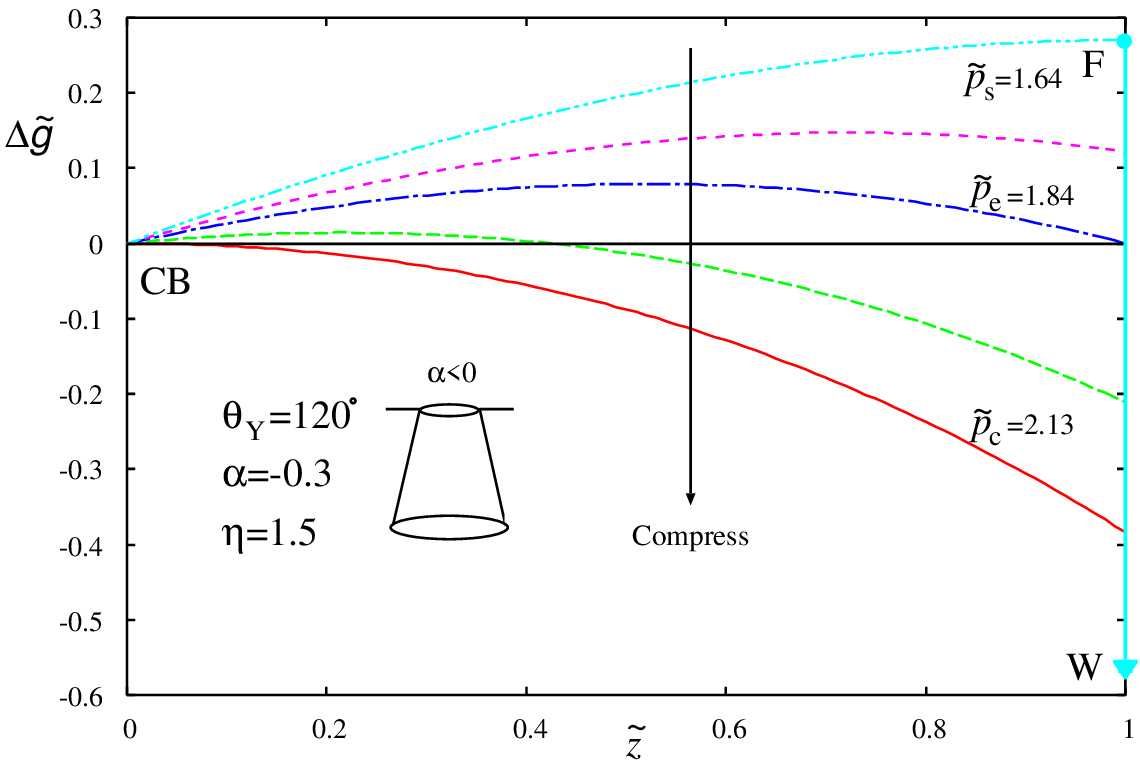}
\label{fig:6a}
}
\subfigure[]
{
\includegraphics[width=0.8\linewidth]{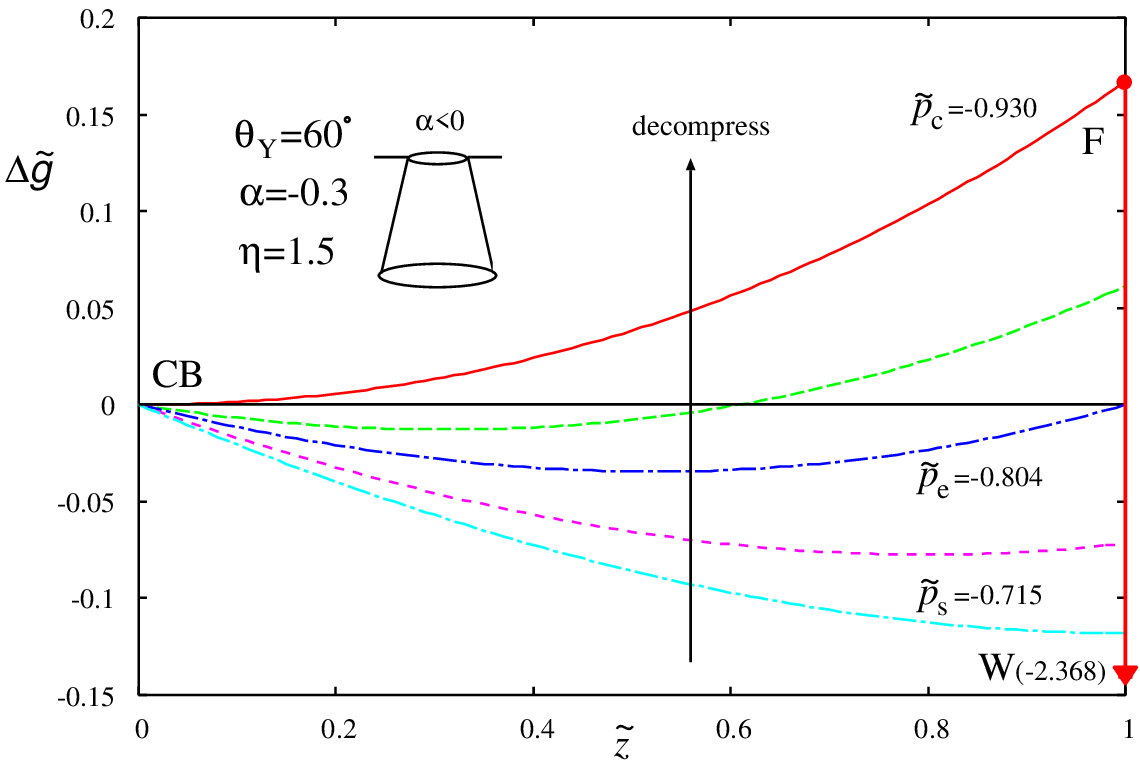}
\label{fig:6b}
}
\end{center}
\caption{
The free-energy landscape between $\tilde{z}=0$ (CB state) and $\tilde{z}=1$ (F state) of the inverted truncated cone with $\alpha=-0.3$ and $\eta=1.5$ when the substrate is (a) hydrophobic ($\theta_{\rm Y}=120^{\circ}>\theta_{\rm c}=78.7^{\circ}$) and (b) hydrophilic ($\theta_{\rm Y}=60^{\circ}<\theta_{\rm c}=78.7^{\circ}$).  The adsorption energy in (a) is $\delta g_{\rm W}=-0.845$ and in (b) is $\delta g_{\rm W}=-2.535$.  Therefore, the free energy of the W state $\Delta g_{\rm W}$ is always lower than that of the F state $\Delta g_{\rm F}=\Delta g\left(\tilde{z}=1\right)$.  In contrast to Fig.~\ref{fig:4}, they are beyond the scale of the figure.  In (a) the applied pressure is $\tilde{p}=\tilde{p}_{\rm s} (1.64), 1.75, \tilde{p}_{\rm e} (1.84), 2.00, \tilde{p}_{\rm c} (2.13)$, and  in (b) the applied pressure is $\tilde{p}_{\rm c} (-0.930), -0.850, \tilde{p}_{\rm e} (-0.804), -0.750, \tilde{p}=\tilde{p}_{\rm c} (-0.715)$ from the top to the bottom. 
 } 
\label{fig:6}
\end{figure}

Figure~\ref{fig:5} shows the three pressures $\tilde{p}_{\rm c}$, $\tilde{p}_{\rm e}$, and $\tilde{p}_{\rm s}$ (Eqs.~(\ref{eq:x24}), (\ref{eq:x34}) and (\ref{eq:x36})), which characterize the free-energy landscape, for the truncated cone with $\alpha=0.3$ (Fig.~\ref{fig:5a}) and the inverted truncated cone with $\alpha=-0.3$ (Fig.~\ref{fig:5b}). These pressures are positive when $\theta_{\rm Y}>\theta_{\rm c}$ and negative when $\theta_{\rm Y}<\theta_{\rm c}$.  The former condition corresponds to the pore under the compressed liquid, and the latter condition corresponds to the pore under the decompressed liquid.  The orders of these three pressures change at $\theta_{\rm Y}=\theta_{\rm c}$, which lead to the inverted free-energy landscape shown in Fig.~\ref{fig:4a} and \ref{fig:4b}.  Moreover, the orders for the narrowing pore ($\alpha=0.3$) and that for the widening pore $\alpha=-0.3$ are reversed, which will further lead to the inverted free-energy landscape for the inverted truncated cone, as shown in Fig.~\ref{fig:6}.  In addition, figure~\ref{fig:5} shows the pressure $\tilde{p}_{\rm CB-W}$ (Eq.~(\ref{eq:x35})), where the free energies of the CB and the W states becomes equal.

Figure~\ref{fig:6a} shows the free-energy landscape for the ink-bottle shaped inverted truncated conical pore with a widening radius of $\alpha=-0.3$ and $\eta=1.5$.  The pore wall is made up of a non-wettable substrate with $\theta_{\rm Y}=120^{\circ}>\theta_{\rm c}=78.7^{\circ}$. In this case, the critical Young's contact angle is $\theta_{\rm c}=78.7^{\circ}$, therefore, the hydrophilic pore ($\theta_{\rm Y}<90^{\circ}$) can sustain the CB state under the positive pressure $\tilde{p}>0$ as far as $\theta_{\rm Y}>\theta_{\rm c}$. The evolution of convex free-energy landscape in Fig.~\ref{fig:6a} is completely different from concave one in Fig.~\ref{fig:4a}; however, it is similar to the free-energy landscape in Fig. ~\ref{fig:4b}. The orders of the three pressures  $\tilde{p}_{\rm c}$, $\tilde{p}_{\rm e}$, and $\tilde{p}_{\rm s}$ are reversed (see also Fig.~\ref{fig:5}, upper tree lines).  The free energy of the W state $\Delta g_{\rm W}$ is lower than that of the F state $\Delta g_{\rm F}=\Delta g\left(\tilde{z}=1\right)$ by $\delta g_{\rm W}=-0.845$ (Eq.~(\ref{eq:x22})), which is almost the entire energy scale of Fig.~\ref{fig:6a}.

When the pressure $\tilde{p}$ increases, the unstable filled F state at $\tilde{z}=1$ becomes metastable at $\tilde{p}_{\rm s}=1.64$ while the CB state remains stable relative to the F state.  In fact, the W state has a much lower free energy $\Delta g_{\rm W}=\Delta g_{\rm F}+\delta g_{\rm W}=0.270-0.845=-0.575$ as compared to that of the CB state.  As a result, the W sate is absolutely stable and the CB state is metastable. As the pressure is further increased, the free energy of the F state decreases and the free-energy barrier develops between the CB and the F states.  When the applied pressure reaches $\tilde{p}_{\rm e}=1.84$, the free energies of the completely filled F and completely empty CB states become equal.  However, the free-energy barrier exists between the CB and the F states.  The magnitude of the free-energy barrier at $\tilde{p}_{\rm e}=1.84$ in Fig.~\ref{fig:6a} and that at $\tilde{p}_{\rm e}=-2.48$ in Fig.~\ref{fig:4b} are on the same order of magnitude.  Therefore, the CB state would remain stable compared to the F state as long as the free-energy barrier remains larger than the thermal energy.

Finally, at the critical pressure $\tilde{p}_{\rm c}=2.13$, the CB state becomes completely unstable and the F state is realized.  Subsequently, the bottom surface becomes wet spontaneously and the completely filled W state is realized.  In contrast to the narrowing pore in Fig.~\ref{fig:4a} where the CB state collapses continuously, the collapse would occur abruptly at $\tilde{p}_{\rm c}$ in the widening pore in Fig.~\ref{fig:6a}.  The prediction of the critical pressure $\tilde{p}_{\rm c}$ in Eq.~(\ref{eq:x25}) is correct in this widening pore.

Figure~\ref{fig:6b} shows the free-energy landscape for the wettable substrate with $\theta_{\rm Y}(=60^{\circ})<\theta_{\rm c}(=78.7^{\circ})$, $\alpha=-0.3$ and $\eta=1.5$. The critical pressure $\tilde{p}_{\rm c}=-0.93$ and it is negative; hence, the liquid must be decompressed and metastable~\cite{Debenedetti1996,Kelton2010}.  The evolution of the free-energy landscape in Fig.~\ref{fig:6b} is very similar to that in Fig.~\ref{fig:4a}.  The free energy of the W state $\Delta g_{\rm W}$ is lower than that of the F state $\Delta g_{\rm F}$ by $\delta g_{\rm W}=-2.535$ (from Eq.~(\ref{eq:x22})), which is beyond the scale of Fig.~\ref{fig:6b}.

Since the substrate is wettable, the W and F states are more stable than the CB state at a positive pressure.  As the liquid is decompressed to a negative pressure, the F state becomes metastable at $\tilde{p}_{\rm s}=-0.715$.  As the liquid is further decompressed, a free-energy minimum appears between the F state at $\tilde{z}=1$ and the CB state at $\tilde{z}=0$, which moves toward the CB state.  Finally, when the pressure $\tilde{p}_{\rm c}=-0.930$, the local minimum reaches the top of the pore at $\tilde{z}=0$.  Although the F state has a higher energy as compared to that of the CB state, the free energy of the W state is $\Delta g_{\rm W}=\Delta g_{\rm F}+\delta g_{\rm W}=0.167-2.535=-2.368$; hence, the W state is more stable than the CB state.  The CB state cannot be recovered simply by deep decompression.  In fact, the adsorption energy $\delta g_{\rm W}$ always acts as the free energy barrier to prevent the W to F dewetting transition.

\subsection{Discussion}

When the size of the pore is in nanoscale, the line tension in addition to the surface tension might play some role in the evolution of free-energy landscape~\cite{Schimmele2007,Guillemont2012,Bormashenko2013,Iwamatsu2016,Law2017}; therefore,  the line tension $\tau$ need to be included in the free energy, which is written as
\begin{equation}
F=\gamma_{\rm lv}S_{\rm lv}-\gamma_{\rm lv}\cos\theta_{\rm Y}S_{\rm sl} + \tau L_{\rm slv},
\label{eq:x37}
\end{equation}
instead of Eq.~(\ref{eq:x2}), where $L_{\rm slv}$ is the perimeter of the three-phase contact line.  In our truncated conical pore with a narrowing radius, for example, the last term of Eq.~(\ref{eq:x37}) becomes
\begin{equation}
\tau L_{\rm slv} = \tau 2\pi f(z) =2\pi\tau R_{0}\left(1-\alpha\tilde{z}\right)
\label{eq:x38}
\end{equation}
which gives a correction to the first order term of $\tilde{z}$ in the scaled free energy in Eq.~(\ref{eq:x10}), and Eq.~(\ref{eq:x17}) is modified to
\begin{equation}
g_{1} = -\left(2\alpha\left(1+\tilde{\tau}\right) + 2\cos\theta_{\rm Y}\sqrt{\eta^{2}+\alpha^{2}} + \tilde{p} \right)
\label{eq:x39}
\end{equation}
with
\begin{equation}
\tilde{\tau}=\frac{\tau}{\gamma_{\rm lv}R_{0}}.
\label{eq:x40}
\end{equation}
If $\tilde{\tau}\ll 1$, the effect of line tension can be safely neglected.  Suppose $R_{0}=10$nm, $\gamma_{\rm lv}=73$mJm$^{-1}$ (water), and $\tau=10^{-11}$N (typical magnitude), we find $\tilde{\tau}\sim 0.014$.  Therefore, this effect can be safely neglected.  Even when the effect of line tension cannot be neglected, the line tension only affects the critical pressure in Eq.~(\ref{eq:x24}) from Eq.~(\ref{eq:x39}).  The line tension will not change the main characteristics of the free-energy landscape $\Delta g\left(\tilde{z}\right)$ because it will remain a cubic polynomial of $\tilde{z}$.

\begin{figure}[htbp]
%Fig.7
\begin{center}
\includegraphics[width=0.65\linewidth]{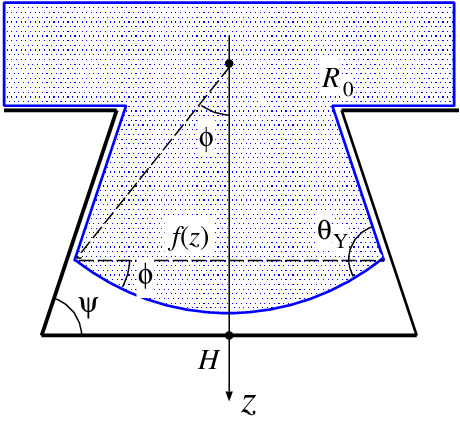}
\caption{
Correction due to the spherical liquid-vapor surface, which contributes to the volume $V$ and the surface area $S_{\rm lv}$.  However, the free energy remains a cubic polynomial of $\tilde{z}$
 }
\label{fig:7}
\end{center}
\end{figure}

In our model we assumed a flat liquid-vapor interface~\cite{Nosonovsky2007,Bormashenko2013}; however, it is possible to consider a spherical surface.  The volume collection $\Delta V$  due to the truncated sphere shown in Fig.~\ref{fig:7} is given by 
\begin{equation}
\Delta V=\frac{\pi}{3}f\left(z\right)^{3}\frac{\left(1-\cos\phi\right)^{2}\left(2+\cos\phi\right)}{\sin^{3}\phi},
\label{eq:x41}
\end{equation}
and the liquid-vapor surface area will be modified such that
\begin{equation}
S_{\rm lv}=2\pi f\left(z\right)^{2}\frac{1-\cos\phi}{\sin^{2}\phi},
\label{eq:x42}
\end{equation}
where $\phi$ is the angle between the liquid-vapor meniscus and the horizontal plane (Fig.~\ref{fig:7}).  It is related to Young's contact angle $\theta_{\rm Y}$ through
\begin{equation}
\phi=\theta_{\rm Y}-\psi,
\label{eq:x43}
\end{equation}
where $\psi$ is the angle of the bottom corner of the pore (Fig.~\ref{fig:7}).  Even with the two corrections in Eqs.~(\ref{eq:x41}) and (\ref{eq:x42}), the free energy is still described by a cubic polynomial of $\tilde{z}$.  The qualitative feature of the free-energy landscapes in Figs.~\ref{fig:4a}, \ref{fig:4b} and \ref{fig:6a}, \ref{fig:6b} will not change.

When the size of the pore is truly nanoscale, the disjoining pressure (DJP) or the surface potential has to be included~\cite{Bormashenko2013}.  Then, we need to determine the liquid-vapor meniscus by solving the Euler-Lagrange equation~\cite{Checco2014}.  A more microscopic approach, such as molecular simulations~\cite{Tinti2017,Amabili2016,Prakash2016,Giacomello2012} or microscopic density functional calculation~\cite{Singh2015,Giacomello2019} would also be necessary.  Certainly, not only quantitative but also a qualitative difference between the microscopic calculations and our macroscopic model would emerge.  For example, due to the DJP, the atomic-scale wetting film always exists~\cite{Evans1985, Iwamatsu1996, Bormashenko2013b}; hence, the apparent pore space would be narrower.  There is also a possibility to observe heterogeneous nucleation of droplets~\cite{Giacomello2012,Jones2017} and bubbles~\cite{Prakash2016} at the bottom corner of the pore.  The singularity of the W state and the discontinuous transition from the F state to W state would be naturally resolved by the effect of the DJP.  These topics are certainly important when the pore size is truly nanoscale and the pressure is high; however, they are beyond the scope of our macroscopic capillary model.

\begin{figure}[htbp]
%Fig.8
\begin{center}
\includegraphics[width=0.85\linewidth]{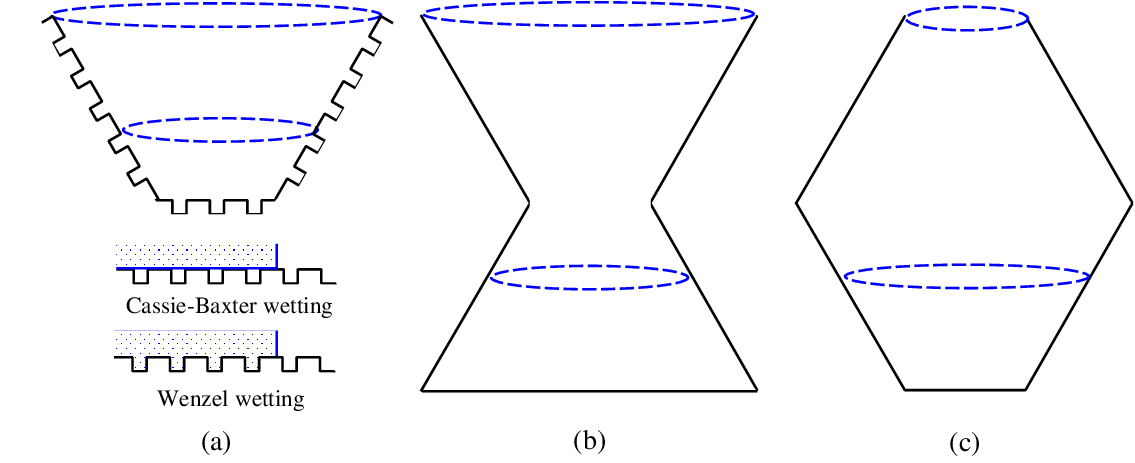}
\caption{
(a) Truncated cylindrical pore with rough inner wall with the microscopic Cassieand the Wenzel wettings. (b) and (c) Two re-entrant structures made from two truncated cylindrical pores.
 }
\label{fig:8}
\end{center}
\end{figure}

Finally, we consider the extension of our simplest conical pore models briefly.  To achieve SHP state, it is well recognized that the hierarchical structure, which increases the solid-liquid surface area, would be advantageous when the substrate is hydrophobic~\cite{Nosonovsky2007,Jiang2020,Bormashenko2013,Giacomello2016,Giacomello2019}.  It is possible to extend our simplest model and introduce sub-structures~\cite{Giacomello2019} as shown in Fig.~\ref{fig:8}(a).  For instance, the wetting of these microscopic structures is modeled by the microscopic wetting models~\cite{Cassie1944,Wenzel1936} (Fig.~\ref{fig:8}(a)) with the apparent averaged contact angles.  If the vapor is trapped in those substructures and we can average the contact angle over the rough surface,  the Young's contact angle $\theta_{\rm Y}$ should be replaced by the apparent contact angle $\theta_{\rm CB}$ given by
\begin{equation}
\cos\theta_{\rm CB}=-1+f_{\rm s}\left(\cos\theta_{\rm Y}+1\right)
\label{eq:x44}
\end{equation}
of the traditional Cassie-Baxter (CB) model~\cite{Cassie1944,Bormashenko2015,Giacomello2016}, where $f_{\rm s}<1$ is the fraction of wet surface area.  Alternatively, we assume that the liquid fill these substructures~\cite{Whyman2011}.  The Young's contact angle $\theta_{\rm Y}$ should be replaced by the apparent contact angle $\theta_{\rm W}$ given by 
\begin{equation}
\cos\theta_{\rm W}=r\cos\theta_{\rm Y}
\label{eq:x45}
\end{equation}
of the traditional Wenzel (W) model~\cite{Wenzel1936,Bormashenko2015,Quere2002}, where $r>1$ is the so-called roughness factor. Then, Young's contact angle $\theta_{\rm Y}$ of the previous subsections must be replaced by either $\theta_{\rm CB}$ or $\theta_{\rm W}$.  Indeed, it is possible to
consider a microscopic intermediate wetting state between
the CB and the W models~\cite{Bormashenko2015}.

\begin{figure}[htbp]
%Fig.9
\begin{center}
\includegraphics[width=0.85\linewidth]{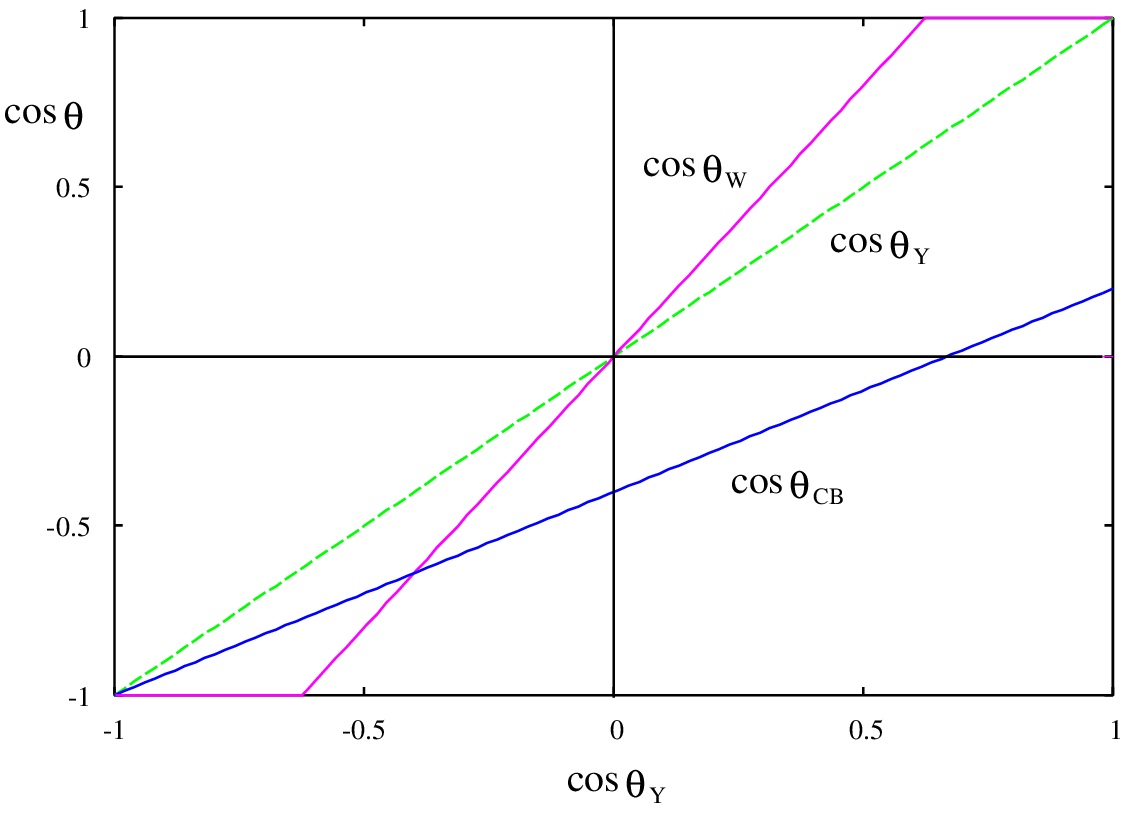}
\caption{
The cosine of the apparent contact angle $\theta_{\rm CB}$ of the Cassie-Baxter model given by Eq.~(\ref{eq:x44}) when $f_{\rm s}=0.6$ and $\theta_{\rm W}$ of the Wenzel model given by Eq.~(\ref{eq:x45}) when $r=1.6$.
 }
\label{fig:9}
\end{center}
\end{figure}

Therefore, an inherently hydrophobic pore wall with $\theta_{\rm Y}>90^{\circ} (\cos\theta_{\rm Y}<0)$ becomes more hydrophobic with $\cos\theta_{\rm W}<\cos\theta_{\rm Y}<0$~\cite{Bormashenko2015,Quere2002} and $\cos_{\rm CB}<\cos\theta_{\rm Y}<0$~\cite{Bormashenko2015} (Fig.~\ref{fig:9}). Both the CB and the Wenzel models of the micro-structures make the pore wall more hydrophobic; therefore, the models are advantageous for the stability of the SHP-CB state. In contrast, when the substrate is hydrophilic ($\cos\theta_{\rm Y}>0$), only the CB model of the pore wall makes the apparent contact angle more hydrophobic ($\cos_{\rm CB}<0$); therefore, the stability of the SHP-CB state will be enhanced. Thus, the roughness of the pore wall or the so-called hierarchical structure could be advantageous to achieve the SHP-CB substrate particularly when the wall is intrinsically hydrophobic.

Of course, the prediction of the W and CB model needs to be considered carefully since the apparent contact angels in Eqs.~\ref{eq:x44} and \ref{eq:x45} of the W and the CB models are the averaged angle which does not take into account the micro structure of the roughness. In fact, it is well recognized that the hydrophobic substrate from hydrophilic materials requires roughness with re-entrant structure~\cite{Joly2009,Amabili2016}.  The Wenzel and the Cassie-Baxter models are macroscopic model so that the range of applicability is certainly limited.

It is also possible to consider the reentrant structures shown in Fig.~\ref{fig:8}(b) and ~\ref{fig:8}(c).  The evolution of the free-energy landscape will be the combination of that of the interface-trapping in Figs.~\ref{fig:4a} and \ref{fig:6b} and that of the barrier-crossing in Figs.~\ref{fig:4b} and \ref{fig:6a}.  For example, the free-energy landscape for the hydrophobic re-entrant structure in Fig.~\ref{fig:8}(b) would be the combination of Fig.~{\ref{fig:4a} and Fig.~\ref{fig:6a}.  Therefore, as the liquid intrudes into the pore, the liquid-vapor interface advances by the interface-trapping followed by the barrier-crossing.  However, for the hydrophobic re-entrant structure in Fig.~\ref{fig:8}(c), the order of the free-energy landscape changes to that in Fig.~\ref{fig:6a} followed by Fig.~\ref{fig:4a}.

Similarly, the free-energy landscape for the hydrophilic re-entrant structure in Fig.~\ref{fig:8}(b) would  be the  combination of Fig.~\ref{fig:4b} and Fig.~\ref{fig:6b}.  On the other hand, the free-energy landscape for the hydrophilic reentrant structure in Fig.~\ref{fig:8}(c) would be the combination of Fig.~\ref{fig:6b} and Fig.~\ref{fig:4b}.  Therefore, more complex free-energy landscape will be expected for the re-entrant
structure whose components are simple truncated cones.

\section{\label{sec:sec4} Conclusion}

In this study, we considered the intrusion and extrusion of liquid into a truncated conical pore~\cite{Jiang2020,Bormashenko2013,Jones2017} and an inverted truncated conical pore~\cite{Jiang2020,Cao2007,Giacomello2019}. The intrusion of liquid corresponds to the destruction of the superhydrophobic Cassie Baxter (SHP-CB) state, and the extrusion of liquid corresponds to the destruction of the Wenzel (W) state and the recovery of the SHP-CB state. We found that the simple criterion of the stability of CB state based on the classical Laplace pressure cannot describe the details of these processes.  

In the truncated conical pore with hydrophobic substrate under the compressed liquid, the destruction of SHP-CB occurs gradually by the movement of the liquid-vapor interface as the liquid is pressurized.  When the substrate is hydrophilic, and the liquid is decompressed, the recovery of the SHP-CB cannot occur since the adsorption energy of liquid on the bottom wall of the pore always prevent the dewetting. 

In contrast, in the inverted truncated conical pore with hydrophobic substrate under the compressed liquid, the destruction of SHP-CB state occurs abruptly since the free-energy barrier exists.  However, the recovery of SHP-CB state under the decompressed liquid cannot occur as the adsorption energy always prevent the dewetting.  Therefore, the geometry of the pore will strongly influence the intrusion and extrusion processes of the liquid in the pore.  These knowledge would be useful to understand and design functional superhydrophobic substrates.

\begin{acknowledgments}
The author is grateful to the reviewers for their insightful commens and useful suggestions, in particular, about the singularity of the Wenzel state of this model. Their criticisms were helpful to shape the final version of this article.
\end{acknowledgments}

%\appendix*

%\section{}
. 

%\nocite{*}
%\bibliography{aipsamp}% Produces the bibliography via BibTeX.

\end{document}